# Comparación de tres técnicas distintas con datos reales de pozo, en la determinación de la permeabilidad*

*Julio Torres*[1,2], *Nuri Hurtado*[1**] *y Milagrosa Aldana*[3]

[1]*Laboratorio de Física Teórica de Sólido, Escuela de Física, Universidad Central de Venezuela (UCV), Caracas 1041A, Venezuela.* [2]*Dpto. de Ciencias Básicas, Universidad Nacional Experimental Politécnica "Antonio José de Sucre" Vicerrectorado "Luís Caballero Mejías", La Yaguara, 1020, Venezuela.* [3]*Dpto. de Ciencias de la Tierra, Universidad Simón Bolívar (USB) Caracas, 1080, Venezuela.*

## Resumen

En el presente trabajo utilizamos tres técnicas distintas para hacer predicciones de permeabilidad en el pozo PX12 del Lago de Maracaibo a partir de datos de porosidad. Una de estas técnicas tiene su base en Lógica Difusa; otra, desarrollada por H. Pape et al, está basada en Teoría Fractal y en las ecuaciones de Kozeny-Carman, y la última técnica es el modelo empírico obtenido por Tixier en 1949. Con el 100% de los datos de porosidad-permeabilidad obtuvimos ecuaciones de predicción para cada técnica. Al comparar los resultados para este porcentaje, encontramos que con el modelo estadístico se obtiene la ecuación que mejor predice el comportamiento de la permeabilidad en profundidad y que los modelos de teoría fractal y empírico son similares. Posteriormente tomamos de forma aleatoria el 25% de los datos para determinar nuevas ecuaciones de predicción. En este caso observamos que los resultados obtenidos por el modelo de teoría fractal son mejores que los obtenidos por cualquiera de los otros dos métodos.

**Palabras clave:**   Fractal; lógica difusa; permeabilidad; porosidad; Tixier.

# Comparison between three different methods with real well data for permeability determination

## Abstract

We have used three different techniques for permeability prediction from porosity data in well PX12 at El Lago de Maracaibo. One of these techniques is statistical and is based in Fuzzy Logic. Another has been developed by H. Pape et al, based in Fractal Theory and the Kozeny-Carman equations, and the other one is an empirical model obtained in 1949 by Tixier. We have used 100% of the permeability-porosity data to obtain the predictor equations in each case. We have found better results with the statistical approach. The results obtained from the fractal model and the Tixier equations are similar in this case. We have also taken randomly 25% of the data to obtain the predictor equations. In this case the best results are those obtained with fractal theory.

## Introducción

A comienzos del siglo 20, se desarrollaron técnicas empíricas basadas en el análisis de registros de pozos para la predicción de parámetros petrofísicos tales como: Permeabilidad, Porosidad, Saturación de Agua, Presión Capilar, etc., los cuales rigen el movimiento de los fluidos (1, 2). Para la segunda mitad del siglo se diseñó un conjunto de técnicas de predicción teórica con las que se pueden determinar estos parámetros petrofísicos. Una de estas técnicas se fundamenta en la ecuación modificada de Kozeny-Carman (3) y en Teoría Fractal, en la que se determina la permeabilidad "k", en función de la porosidad "$\phi$", el exponente de cementación "m" y la dimensión fractal "D" (3, 4). Esta técnica hace uso de la relación que tiene la permeabilidad con la geometría del espacio poroso y no con las propiedades de los fluidos. Otra de las técnicas utiliza los conceptos de Lógica Difusa, basada en los trabajos desarrollados por Lofti Zathe, Mamdani, Assiliam, Takagi, Sugeno y Kang (5, 6). En esta técnica se define un grupo de funciones de membresía que son una representación matemática del grado de pertenencia de los datos de entrada respecto al conjunto de salida, determinando así una respuesta (6). La relación entre las variables se representa por implicaciones If, Then (If Hipótesis; Then-Conclusión), donde Hipótesis son los antecedentes y Conclusión las consecuencias, es decir, si un hecho es conocido entonces otro puede ser inferido.

En este trabajo hicimos un estudio comparativo de los resultados obtenidos usando la ecuación empírica desarrollada por Tixier, el modelo basado en Teoría Fractal y un modelo basado en Lógica Difusa, para predecir la permeabilidad en el pozo Px12 del Bloque III Lago de Maracaibo.

## Modelos

### Modelo empírico de Tixier

En 1949 Tixier (2), usando relaciones empíricas entre la saturación de agua, la resistividad y la presión capilar, estableció un método para determinar la permeabilidad a través de los gradientes de resistividad, que relacionan la permeabilidad con la porosidad y la saturación de agua irreducible como se muestra en la ecuación [1]:

$$k^{\frac{1}{2}} = 250 \frac{\phi^3}{S_{wi}} \qquad [1]$$

### Modelo basado en Teoría Fractal:

Mediante la ecuación de Kozeny-Carman (3) se obtiene una ecuación generalizada para determinar la permeabilidad en función de la porosidad "$\phi$" (ecuación [2]), el exponente de cementación "m" $(Exp_1 = m)$, la dimensión fractal "D"

$$\left(D = 3 + \frac{\log(\phi/0,534)}{0,391 * \log(1/2\phi)^4}\right); Exp_2 = m + \frac{2}{c_1(3-D)}$$

y una constante empírica que "$c_1$" ($c_1 = 0,263\phi^{-0,2}; 0,39 < c_1 < 1$) depende de la porosidad (3):

$$k = a\phi + b\phi^{Exp_1} + c(10\phi)^{Exp_2} \qquad [2]$$

donde los parámetros $a$, $b$ y $c$ dependen del área de estudio y deben ser determinados. En este modelo, el radio de poro efectivo y la permeabilidad pueden ser determinados suponiendo un estructura multifractal.

### Modelo basado en Lógica Difusa

A diferencia de la lógica Booleana, la lógica difusa permite asignar a un cierto elemento un grado de pertenencia o membresía (7), respecto a un conjunto dado; el grado de membresía es un número real entre cero (0) y uno (1) y la curva que describe el paso entre estos dos valores se denomina función de membresía. En este trabajo usamos, para predecir permeabilidad, el modelo de Takagi-Sugeno-Kang (TSK) (5, 6), el cual nos permite considerar como parte de la entrada una función lineal o constante, lo que facilita el cómputo de los valores de salida. Para entrenar este modelo difuso, se usó como variable de entrada el valor de porosidad de núcleo (fracción) y como variable de salida el logaritmo de la permeabilidad.

## Resultados

Para determinar las funciones de predicción de permeabilidad con cada uno de los modelos antes señalados, se hicieron dos tipos de pruebas. Primero se tomó el 100% de los datos de núcleo del pozo en estudio, obteniéndose las ecuaciones de predicción de permeabilidad y el factor de correlación. En la segunda parte, se escogió de forma aleatoria el 25% de los datos de núcleo, con los que se determinaron nuevas ecuaciones de predicción de permeabilidad y su correspondiente factor de correlación.

En la Figura 1 se muestran los datos de permeabilidad medida en núcleo, así como la permeabilidad predicha por el modelo difuso con el 100% de los datos, en función de la profundidad. En la gráfica se puede observar que con esta técnica se ha logrado predecir el comportamiento cualitativo de los datos de permeabilidad, especialmente en la zona de menor profundidad. Esto probablemente se deba a que en la primera zona el pozo está compuesto por arenas más limpias. Los resultados con teoría fractal para el 100% de los datos no mejoran los obtenidos con Lógica Difusa, lo cual se refleja en un factor de correlación menor. El modelo empírico de Tixier arroja resultados similares a los obtenidos con el modelo fractal.

En la Figura 2 graficamos la predicción de permeabilidad obtenida con el modelo de teoría fractal, para el 25% de los datos escogidos al azar, así como los datos medidos en núcleo. Se puede observa que la predicción es bastante buena en todo el rango de profundidad que estamos estudiando. Sin embargo, al igual que en la Figura 1, se encuentra una mejor predicción a menor profundidad. Los resultados obtenidos para este porcentaje y con las otras técnicas, no mejoran los derivados del modelo Fractal.

En la Tabla 1 presentamos los valores de los parámetros encontrados con un ajuste polinomial, para determinar la ecuación de permeabilidad del modelo Fractal, el tipo y número de funciones de membresía de salida óptimas para la determinación de la permeabilidad del modelo Difuso y el factor de correlación para cada modelo. Los resultados indican que el modelo estadístico permite una mejor aproximación al tener una mayor densidad de datos, tal como se espera.

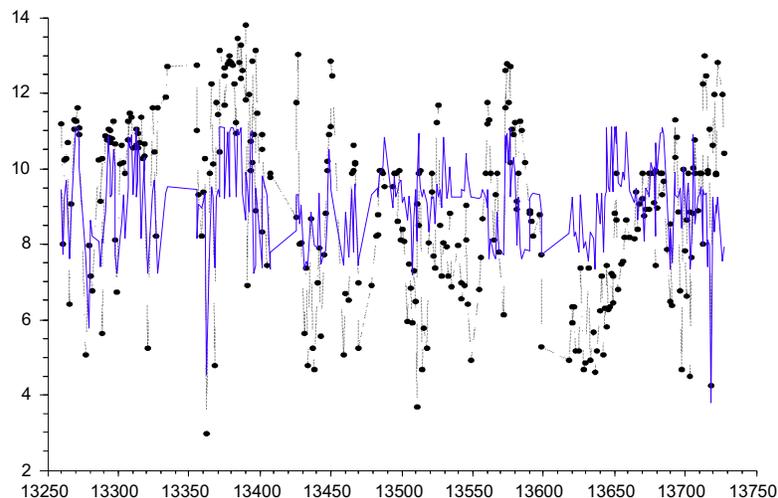

Figura 1.　Representación de la permeabilidad medida en núcleo (puntos y línea punteada) y permeabilidad predicha con el modelo Difuso usando 100% de los datos (línea continua), en función de la profundidad.

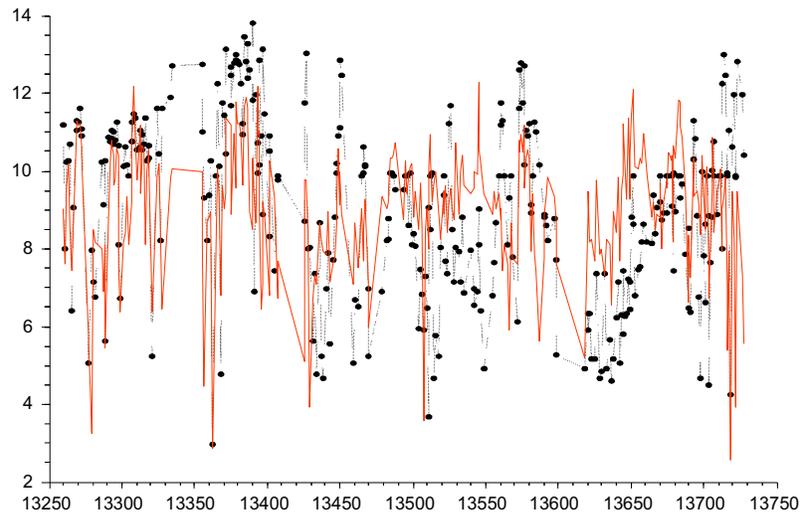

Figura 2. Permeabilidad medida en núcleo (puntos y línea punteada) y permeabilidad predicha con el modelo Fractal usando el 25% de los datos escogidos al azar (línea continua), en función de la profundidad.

Tabla 1
Parámetros y funciones determinados tanto para el modelo fractal, como para el de lógica difusa, con los distintos % de datos de entrada. En cada modelo y para cada % se muestra el factor de correlación "r"

|  | Teoría Fractal | | | | Lógica Difusa, Funciones | | | Tixier |
|---|---|---|---|---|---|---|---|---|
| Datos | a | b | c | r | Tipo | N° | r | r |
| 100% | 539,7 | 11240 | 868,7 | 0,429 | Triangular | 6 | 0,497 | 0,434 |
| 25% al azar | -379,5 | 13400 | 542,2 | 0,443 | Triangular | 6 | 0,316 | |

Comparando los resultados obtenidos por las técnicas al variar el porcentaje de datos de entrada, se puede concluir que la predicción de permeabilidad usando el modelo fractal da mejores resultados con la disminución de la cantidad de datos de núcleo, como se muestra en la Tabla 1. Este resultado es razonable ya que el modelo de Kozeny-Carman (KC) fue creado suponiendo condiciones muy ideales de poros y cuellos sin ningún tipo de rugosidades (3). Así, al disminuir la densidad de información, podríamos estar acercándonos más al modelo de KC ideal. Esto podría explicar los resultados obtenidos con el 25% de los datos al azar, donde el modelo de teoría fractal da una mejor predicción que el de lógica difusa y que el modelo de Tixier para el 100%.

## Conclusiones

En este trabajo realizamos un estudio comparativo entre tres modelos para predecir permeabilidad en el pozo PX12 del Bloque III Lago de Maracaibo: la ecuación empírica desarrollada por Tixier, un modelo basado en Teoría Fractal y otro modelo basado en Lógica Difusa.

Usando el 100% de los datos de porosidad-permeabilidad para determinar las ecuaciones de predicción de cada técnica, encontramos que con el modelo estadístico se obtiene la ecuación que mejor predice el comportamiento de la permeabilidad en profundidad. Los resultados obtenidos por los modelos basados en la teoría fractal y formulaciones empíricas de Tixier dan resultados similares para este porcentaje.

La toma aleatoria del 25% de los datos para determinar nuevas ecuaciones de predicción, indica que los resultados obtenidos con el modelo basado en teoría fractal son mejores que los obtenidos con el modelo difuso, e inclusive mejor que los resultados con el 100% para el modelo de Tixier. Esto podría deberse a que una disminución de la densidad de los datos estaría acercándonos más al modelo de KC (modelo ideal del medio poroso).

## Referencias Bibliográfica

1. NELSON P.H. ***The log Analyst*** 35(3): 38-62, 1994.
2. BALAN B., MOHAGHEGH S., AMERI S. ***SPE30978***: 1–10, 1995.
3. PAPE H., CLAUSER C., IFFLAND J. ***Geophysics*** 64(5): 1447–1460, 1999.
4. PAPE H., CLAUSER C., IFFLAND J. ***Pure and Applied Geophysics*** 157: 603–619, 2000.
5. FINOL J., GUO Y.K., JING X.D. ***Journal of Petroleum Science and Enginiering*** 29: 97–113, 2001.
6. FINOL J., JING X.D. ***Geophysics*** 67(3): 817–829, 2002.
7. WONG K.W., WONG P.M., GEDEON T.D., FUNG C.C. ***APPEA Journal***: 587–593, 2003.